\pdfoutput=1

\documentclass[11pt,twoside,a4paper,cmspaper,final,collab]{cms-tdr}
\def\svnVersion{175948}\def\cmsCernNo{2012-232}\def\cmsCernDate{\today}\def\cmsMessage{Submitted to Physics Letters B}

\begin{document}\cmsNoteHeader{EXO-11-102}

\hyphenation{had-ron-i-za-tion}
\hyphenation{cal-or-i-me-ter}
\hyphenation{de-vices}
\RCS$Revision: 160660 $
\RCS$HeadURL: svn+ssh://svn.cern.ch/reps/tdr2/papers/EXO-11-102/trunk/EXO-11-102.tex $
\RCS$Id: EXO-11-102.tex 160660 2012-12-09 08:51:47Z mmozer $
\newlength\cmsFigWidth
\ifthenelse{\boolean{cms@external}}{\setlength\cmsFigWidth{0.85\columnwidth}}{\setlength\cmsFigWidth{0.38\textwidth}}
\ifthenelse{\boolean{cms@external}}{\providecommand{\cmsLeft}{top}}{\providecommand{\cmsLeft}{left}}
\ifthenelse{\boolean{cms@external}}{\providecommand{\cmsRight}{bottom}}{\providecommand{\cmsRight}{right}}
\cmsNoteHeader{EXO-11-102} 
\renewcommand{\MET}{\ensuremath{{E\!\!\!/}_{\!\mathrm{T}}}\xspace}
\newcommand{\pp}{\Pp\Pp}%
\newcommand{\Wo}{\PW}%
\newcommand{\Wp}{\PWp}%
\newcommand{\Wm}{\PWm}%
\newcommand{\Zo}{\cPZ}%
\newcommand{\Ho}{\PW}%
\newcommand{\ra}{\ensuremath{\rightarrow}}%
\newcommand{\MN}{\Pgm\Pgn}%
\renewcommand{\LL}{\ensuremath{\ell^+\ell^-}}%
\renewcommand{\EE}{\Pep\Pem}%
\newcommand{\EN}{\Pe\cPgn}%
\newcommand{\LN}{\ensuremath{\ell\cPgn}}%
\newcommand{\MW}{\ensuremath{m_\PW}}%
\newcommand{\MZ}{\ensuremath{m_\cPZ}}%
\newcommand{\MT}{\ensuremath{M_\mathrm{T}}}%
\newcommand{\MLL}{\ensuremath{m_{\ell\ell}}}%
\newcommand{\nunubar}{\ensuremath{\cPgn\cPagn}}%
\newcommand{\qqbar}{\ensuremath{\cPq\cPaq}}%
\newcommand{\HZZ}{\ensuremath{\PH\ra\cPZ\cPZ}}%
\newcommand{\HZZllqq}{\ensuremath{\HZZ\ra\qqbar\,\LL}}%
\newcommand{\Hllqq}{\ensuremath{\PH\ra\LL\qqbar}}%
\newcommand{\RedMPl}{\ensuremath{\overline{M}}_{\mathrm{Pl}}}%
\newcommand{\MPl}{\ensuremath{M_{\mathrm{Pl}}}}%
\newcommand{\Grav}{\ensuremath{\mathrm{G}_\mathrm{KK}}}%
\newcommand{\GZZ}{\ensuremath{\Grav\ra\cPZ\cPZ}}%
\newcommand{\GZZllqq}{\ensuremath{\GZZ\ra\qqbar\,\LL}}%
\newcommand{\Gllqq}{\ensuremath{\Grav\ra\LL\qqbar}}%
\newcommand{\XZZ}{\ensuremath{X\ra\cPZ\cPZ}}%
\newcommand{\mZ}{\ensuremath{m_{\cPZ}}}%
\newcommand{\mW}{\ensuremath{m_{\PW}}}%
\newcommand{\mH}{\ensuremath{m_{\PH}}}%
\newcommand{\mZZ}{\ensuremath{m_{\cPZ\cPZ}}}%
\newcommand{\mll}{\ensuremath{m_{\ell\ell}}}%
\newcommand{\mjj}{\ensuremath{m_{jj}}}%
\newcommand{\GeVnn}{\GeV}
\newcommand{\GeVcnn}{\GeV}
\newcommand{\GeVsnn}{\GeV}
\newcommand{\ktilde}{\ensuremath{\tilde{k}}}%
\title{Search for a narrow, spin-2 resonance decaying to a pair of $\cPZ$ bosons in the $\cPq\cPaq\ell^+\ell^-$ final state}

\date{\today}

\abstract{
Results are presented from a search for a narrow, spin-2 resonance decaying
into a pair of $\cPZ$ bosons, with one $\cPZ$-boson decaying into leptons
(\Pep\Pem\ or \Pgmp\Pgmm) and the other into jets.
An example of such a resonance is the Kaluza--Klein graviton, $\Grav$,
predicted in Randall--Sundrum models. The analysis is based on a 4.9\fbinv sample of proton-proton collisions at a center-of-mass energy of
7\TeV, collected with the CMS detector at the LHC.
Kinematic and topological properties including decay angular distributions
are used to discriminate between signal and background. No evidence
for a resonance is observed, and upper limits
on the production cross sections times branching fractions are
set. In two models that predict \cPZ-boson spin correlations in
graviton decays, graviton masses are excluded lower than a value
which varies between 610 and 945\GeV, depending on the model and the
strength of the graviton couplings.
}

\hypersetup{%
pdfauthor={CMS Collaboration},%
pdftitle={Search for a narrow spin-2 resonance decaying to a pair of Z vector bosons in the semileptonic final state},%
pdfsubject={CMS},%
pdfkeywords={CMS, physics, Kaluza-Klein, graviton}}

\maketitle 

\section{Introduction}

The standard model (SM) of particle physics does not include an
explanation of the large difference between the typical scales of
gravity and the other fundamental forces (i.e., the ``hierarchy
problem'').  This problem can be overcome by adding one warped spatial
dimension to space-time, according to the Randall--Sundrum
model~\cite{RS1Graviton1,RS1Graviton2}; in such a scenario, the
existence of Kaluza--Klein (KK) excitations of a spin-2 boson, the
graviton, is predicted. These models have two fundamental parameters:
the mass $M_{1}$ of the first graviton mode, referred to as \Grav\ in
the following, and the dimensionless coupling strength $\ktilde = k /
\RedMPl$, where $k$ is the curvature of the warped space and $\RedMPl$
the reduced Planck mass ($\RedMPl \equiv
\MPl / \sqrt{8 \pi}$).

In the Randall--Sundrum model discussed above (RS1), the production
cross section in pp collisions at $\sqrt{s}=7\TeV$ for a KK graviton of
mass $M_{1}=700\GeV$ and $\ktilde=0.10$ is 22\unit{pb}. The branching fraction to
\cPZ\cPZ~bosons is about 4\%, comparable to those of the diphoton and
dilepton final states. These channels are therefore the most sensitive, and
several searches for such signatures of the
RS1 scenario have been performed at the Tevatron
\cite{CDF:RS1Limit1,CDF:RS1Limit2,D0:RS1Limit1} and at the Large Hadron Collider (LHC)
\cite{CMS:RS1Limit1,CMS:RS1Limit2,ATLAS:RS1Limit1,ATLAS:RS1Limit2}.

However, different models with warped extra dimensions (ED) allow the
SM fields to propagate in the ED
\cite{RSGravAgashe1,RSGravFitz1,RSGravAntipin1}.  In these models, as
a consequence of the localization of SM particles near the Planck or
the TeV brane, decays to diphotons and dileptons are suppressed by a
factor proportional to the volume of the
ED~\cite{RSGravFitz1}.  This scenario is more compatible with
electroweak precision tests and limits on flavor-changing neutral
current processes than the original RS1. The different couplings of
the graviton to the SM fields result in two distinctive effects: the
branching fraction to SM vector-boson pairs can become dominant for
certain values of the model parameters, and a very strong enhancement
in the longitudinal polarization of the vector bosons is predicted.
Because of the aforementioned suppression of photon and fermion
couplings, the total production cross section is also smaller with
respect to RS1 gravitons: in the Agashe--Davoudiasl--Perez--Soni (ADPS)
model~\cite{RSGravAgashe1} for $M_{1}=700\GeV$ and $\ktilde=0.50$, it amounts to 0.31\unit{pb}
in pp collisions at $\sqrt{s}=7\TeV$, and the branching fraction to
longitudinally polarized \cPZ\cPZ~bosons is about 12\%.

The CDF and ATLAS experiments have searched for beyond-SM particles
decaying to $\cPZ\cPZ$ pairs in the $2 \ell 2$q, $4 \ell$ and $2 \ell 2\cPgn$
final states~\cite{CDF:ZZsearch,ATLAS:ZZsearch},
without specific requirements on the spin or the production mechanism
of the particle.  This Letter presents a search for heavy narrow
spin-2 resonances in the $\cPZ\cPZ$ final state and investigates both the
original RS1 model and the ADPS
model.
The distinct angular distributions
resulting from different polarizations in the decay are exploited to
enhance the signal sensitivity.  The process studied, $\GZZllqq$
($\ell = \Pe,\Pgm$), combines a high branching fraction with good mass
resolution and limited background rates.
The dominant background is \Zo-boson production with associated jets
(``Z+jets'').
The cross section for this process falls rapidly with the invariant mass
of the candidate \Zo-boson pair, $\mZZ$.

This analysis closely follows a search for the Higgs boson in the same final
state~\cite{HIG-11-027} regarding lepton and jet selection criteria. However,
a new angular likelihood discriminant is introduced, tuned to appropriate graviton models, which is more sensitive to the ED models under consideration. The search is also extended to a higher mass range.

\section{Event reconstruction and selection}
\label{sec:reco}

We search for a fully reconstructed decay chain of the
graviton $\GZZllqq$, where the charged leptons $\ell^\pm$ are either
both muons or both electrons, and each quark is associated with a jet in the
Compact Muon Solenoid (CMS) detector.
This search covers a graviton mass range
between $400$ and $1200\GeV$. At higher masses, the two jets from the
hadronic Z-boson decay merge into one, therefore the efficiency of the
present search signature is degraded.

A detailed description of the CMS detector can be found in
Ref.~\cite{CMS:2010}.  In the cylindrical coordinate system of CMS,
$\phi$ is the azimuthal angle and the pseudorapidity ($\eta$) is
defined as $\eta = - \ln[\tan(\theta/2)]$, where $\theta$ is the polar
angle with respect to the counterclockwise-beam direction.  The
central feature of the CMS detector is a 3.8\unit{T} superconducting
solenoid of 6\unit{m} internal diameter.  Within the field volume are the
silicon tracker, the crystal electromagnetic calorimeter (ECAL), and
the brass-scintillator hadron calorimeter (HCAL).
The muon detectors are located outside the solenoid and are installed
between the layers of the steel yoke of the flux return.
In addition, CMS has extensive forward calorimetry, in
particular two steel/quartz-fiber hadron forward (HF) calorimeters, which
cover the pseudorapidity range $2.9 < |\eta| < 5.2$.

Although the main sources of background are estimated from data, Monte
Carlo (MC) simulations are used to develop and validate the methods
used in the analysis. Background samples are generated using either
\MADGRAPH 5~\cite{MADGRAPH} (Z+jets, $\ttbar$),
\SHERPA 1.13~\cite{ref:sherpa} (Z+jets for cross-checks and systematic uncertainties),
or \PYTHIA 6.4.24~\cite{PYTHIA} (di-boson backgrounds). Signal events
are generated at Leading Order (LO), for both the RS1 and the ADPS
model, using a dedicated generator described in Ref.~\cite{Gao:2010qx}
and interfaced to \PYTHIA for parton showering and hadronization.  Parton
distribution functions (PDFs) are modeled using the parameterizations
\textsc{ct10} at next-to-LO~\cite{Lai:2010vv} and
\textsc{cteq6}~\cite{CTEQ6:2004} at LO.  For both signal and
background MC, events are simulated using a
\GEANTfour-based model~\cite{GEANT4} of the CMS detector and
processed using the same reconstruction algorithms as for data.

Muons are measured with the silicon tracker and the muon
system~\cite{CMS-PAS-MUO-10-004}.  Electrons are identified as tracks
in the tracker pointing to energy clusters in the
ECAL~\cite{CMS-PAS-EGM-10-004}.

Muon events are triggered by segments in the outer muon system
matching online reconstructed tracks, in a pseudorapidity range
$|\eta|<2.4$. Trigger thresholds on the momentum transverse with respect to
the pp beam direction, \PT, range
from 7 to 17~$\GeVcnn$ for the leading muon and from 7 to 8~$\GeVcnn$
for the subleading muon, depending on the instantaneous
luminosity. The double-muon trigger is supplemented with a single-muon
trigger with a $\PT$ threshold of 24~$\GeVcnn$ and loose isolation
requirements. Electron events are triggered by clusters in the ECAL
matched to online reconstructed tracks, covering a pseudorapidity
range $|\eta|<2.5$ and with $\PT$ thresholds of 17~(8)~$\GeVcnn$ for
the leading (subleading) electron. Loose cluster shape, as well as
calorimetric and track isolation requirements are applied as well,
becoming stricter with increasing instantaneous luminosity. After offline
requirements, the trigger efficiency is 100\% within uncertainties for
electrons and $>95\%$ for muons.

In the offline selection, muons and electrons are required to have \PT
greater than 20\GeVcnn. At least one lepton must have $\PT >
40\GeVcnn{}$.  Leptons are measured in the pseudorapidity range
$|\eta|<2.4$ for muons, and $|\eta|<2.5$ for electrons. Electrons in
the transition range between the barrel and endcap,
$1.44<|\eta|<1.57$, are excluded.  Leptons are required to be isolated
from hadronic activity in the detector using the quantities
$I_{\mathrm{trk}}$, $I_{\mathrm{ECAL}}$ and $I_{\mathrm{HCAL}}$, which
represent respectively the sum of transverse momentum (energy) in the
tracker (ECAL and HCAL), within a surrounding cone of $\Delta R
\equiv \sqrt{(\Delta\eta)^2+(\Delta\phi)^2} <0.3$
excluding the lepton candidate itself, where $\Delta\eta$ and
$\Delta\phi$ are the differences in pseudorapidity and in azimuthal
angle. For muons, the sum $I_{\mathrm{trk}} + I_{\mathrm{ECAL}} + I_{\mathrm{HCAL}}$ is required to be less than
15\% of the measured $\PT$ of the muon. For electrons, the same isolation
requirement
is applied but additional selections on the cluster shape and
track-cluster matching are set to ensure proper identification.
 The details of electron and muon identification criteria are described
in Ref.~\cite{VBTF}.

Jets are reconstructed with the particle flow
(PF)~\cite{CMS-PAS-PFT-09-001} algorithm.
Reconstructed particle candidates are clustered to form jets with the
anti-$k_\mathrm{T}$ algorithm~\cite{Cacciari:2008gp} with a distance parameter of $0.5$.
 Jets are required to be inside the tracker
acceptance ($|\eta|<2.4$) and to have $\PT>30$\GeVcnn.  Jets that
overlap with isolated leptons within $\Delta R=0.5$ are removed.
Jet energy corrections are applied to account for the non-linearity due to
instrumental effects.  These corrections are based on in-situ
measurements using dijet, $\gamma$+jet and Z+jets data
samples~\cite{Chatrchyan:2011ds}. A quality selection, primarily based
on the energy balance between charged and neutral hadrons, is applied
to remove spurious jets due to detector artifacts. The jet four-momentum
is corrected for
the contribution of overlapping minimum bias events coming from
different proton-proton collisions (pileup) based on the effective jet area,
following Ref.~\cite{Cacciari:2008gn}.

Pairs of oppositely charged leptons and jets passing the
above selections are
considered as $\Zo$-boson candidates.  Requirements on the dilepton and dijet
invariant masses, $\mll$ and
$\mjj$ are used to suppress background. The requirement $75<\mjj<105\GeVsnn$
is applied in order to
reduce the $\Zo$+jets background, and $70<\mll<110\GeVsnn$ to reduce
background without a genuine~$\Zo$ decaying to leptons, such as
$\ttbar$.

The kinematics of the jets is corrected by constraining the invariant
mass of the dijet system to the mass of the $\Zo$ boson.
Experimental resolutions, typically lower than 10\%, are taken into
account in the kinematic fit. The
statistical analysis is based on the invariant mass of the graviton
candidate, $\mZZ$, calculated by imposing the kinematic constraint
to the dijet system.

\subsection{Jet flavor analysis}
\label{sec:jet}

As the dominant background is \Zo+jets ($\Zo\rightarrow \LL$)
production, we use the flavor content of
the jets compared to those from \Zo-boson decays to discriminate against
it.  In \Zo+jets events the jets will predominantly be produced by the
fragmentation of low-mass quarks and gluons.
As a result, we can exploit the fact that, in \Zo\Zo~events with one
\Zo-boson decaying into jets, there will be a larger fraction of jets from
heavy-quark fragmentation than from \Zo+jets, and
there will be no jets from gluon hadronization.
We take advantage of both features by tagging the b flavor and introducing a
likelihood discriminant that separates gluon and light-quark jets on a
statistical basis, as described in Ref.~\cite{HIG-11-027}.

To identify jets originating from the hadronization of bottom quarks,
we use the CMS track counting high-efficiency (TCHE)
\cPqb-tagging algorithm~\cite{CMS-PAS-BTV-10-001, CMS-PAS-BTV-11-001}.
The data are split into three \cPqb-tag categories.  The 2
\cPqb-tag category is required to have one jet with a value of
the TCHE discriminant passing the medium working point selection
(${\sim}65\%$ efficiency) and the other jet passing the loose working
point selection (${\sim}80\%$ efficiency). Events not selected in the 2
b-tag region are categorized as 1 b-tag if they have at least one jet
satisfying the loose \cPqb-tagging requirement. The 0 b-tag
category contains all the remaining events.

For the 0 \cPqb-tag category, which is dominated by the $\Zo$+jets
backgrounds, a quark-gluon discriminant ($QG$) is constructed from
observables related to the multiplicity and distribution of the jet
constituents \cite{HIG-11-027}. The $QG$-discriminant values range from 0 (gluon-like) to 1 (quark-like).  We apply a loose requirement, $QG >
0.1$, in the 0 \cPqb-tag category only.

In order to suppress
the substantial $\ttbar$ background in the 2 \cPqb-tag
category, a discriminant, $\lambda$, is constructed using resolution
functions of the PF jets as described in Ref.~\cite{Chatrchyan:2011tn},
to provide a measure of the missing
transverse energy ($\MET$) significance.  We apply a loose
requirement, $2\ln{\lambda} (\MET) < 10$, in the 2 \cPqb-tag
category only.

\subsection{Angular analysis}
\label{sec:anglik}

Due to the tensor nature of the graviton, the angular distribution of
its decay products shows a distinct signature.  Five angles
$(\theta^*, \theta_1, \theta_2, \Phi_1, \Phi)$ defined in
Ref.~\cite{Gao:2010qx}
fully describe the kinematics of the $\qqbar, \Pg\Pg \to\GZZllqq$
process: these angles are only weakly correlated with the invariant masses of
$\Grav$ and the two $\Zo$ bosons, and with kinematic quantities of
the graviton candidate in the laboratory frame.

\begin{figure}[htbp]
\centering
\includegraphics[width=0.45\textwidth]{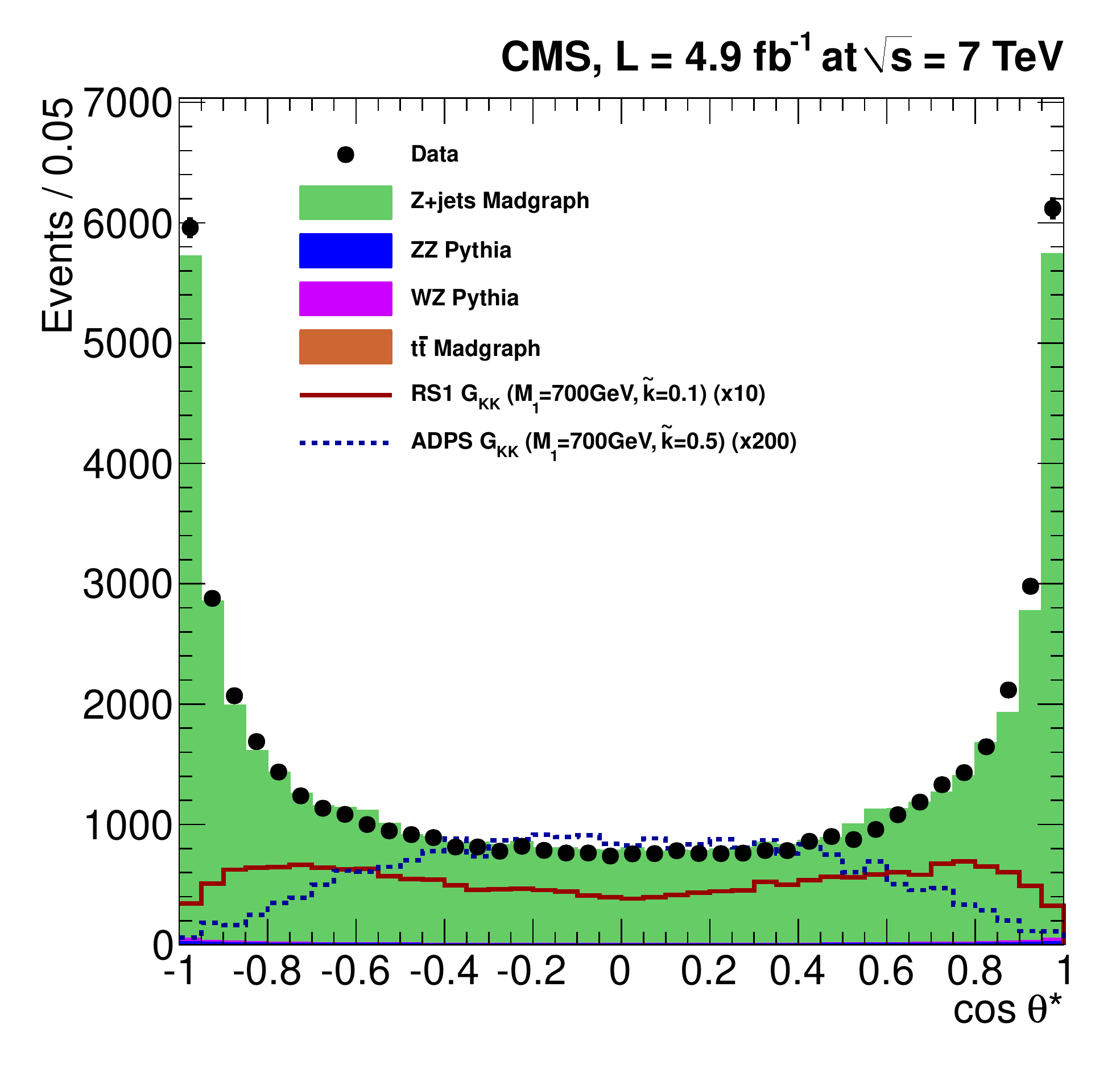}
\includegraphics[width=0.45\textwidth]{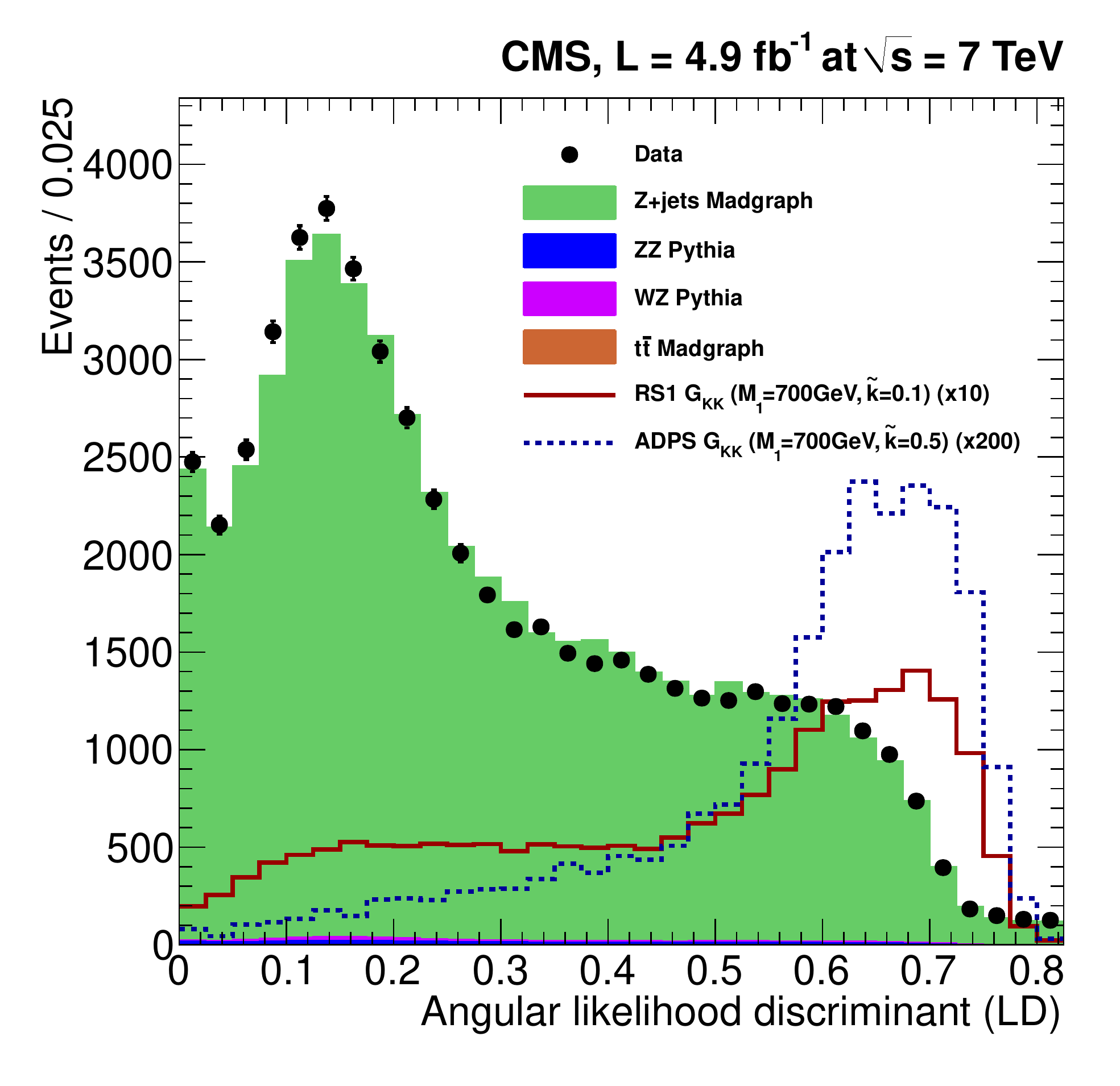}
\caption{
Distributions of the $\cos\theta^*$ variable (\cmsLeft) and the angular
likelihood discriminant, $LD$ (\cmsRight), after pre-selection
requirements.  Points with error bars show the data distributions.
Stacked, filled histograms show the background expectations from
simulated events with the different component breakdown. The
contribution from Z+jets dominates the background. Solid- or
dashed-line histograms (overlaid) show the expected distributions for
RS1 and ADPS gravitons with mass $M_1 = 700\GeV$,
scaled up by arbitrary factors for illustration.}
\label{fig:data_loose}
\end{figure}

An angular likelihood discriminant ($LD$) is defined based on the
probability ratio of the signal and background hypotheses, as
described in Ref.~\cite{Gao:2010qx}. These probabilities are
parameterized as functions of $\mZZ$: the full five-dimensional
theoretical distributions multiplied by acceptance functions are used
for the two signal types, while empirical functions describe the
background probabilities.  The \cmsLeft\ panel of
Figure~\ref{fig:data_loose} shows the distribution of $\cos\theta^*$
after the pre-selection requirements, where $\theta^*$ is the polar
angle of one of the Z boson momentum vectors in the $\cPZ\cPZ$ rest frame,
measured with respect to the beam axis. Because the distributions of
the signal and background events are so different, $\cos\theta^*$ is
an important part of $LD$. The \cmsRight\ panel of
Figure~\ref{fig:data_loose} presents the $LD$ distribution after the
same pre-selection requirements. The expectation of longitudinal
polarization in the ADPS model results in a visible enhancement of the
discriminating power of the $LD$ with respect to the Z+jets
background.

The figure of merit proposed in Ref.~\cite{punzi} is used to optimize
the selection on the angular likelihood discriminant in order to
maximize the expected sensitivity to the production of an RS1 graviton.
Because of the different background contributions,
the optimization is performed independently for the three b-tag categories
of events. The results are approximately
independent of the graviton mass hypothesis and lead to cut values of
$LD>0.28$, 0.35, and 0.21 for the 0 b-tag, 1 b-tag and 2 b-tag categories,
respectively. The optimization is performed only on the RS1 signal
and the same selection criteria are used for both models
under scrutiny.

\begin{table}[hbt]
\begin{center}
\topcaption{Summary of optimized kinematic and topological selection requirements.}
\begin{tabular}{lc|c|c}
\multicolumn{1}{c}{} & \multicolumn{3}{c}{ Pre-selection }\\
\hline
${\PT(\ell^\pm)}$  & \multicolumn{3}{c}{$\PT > 40~(20)\GeVcnn$}\\
${\PT(\text{jets})}$ & \multicolumn{3}{c}{$>30$\GeVcnn} \\
${ |\eta|(\ell^\pm)}$ &  \multicolumn{3}{c}{$\Pe^\pm:\ |\eta|<2.5$,\ $\Pgm^\pm:\ |\eta| <2.4$} \\
${ |\eta|(\text{jets})}$ &  \multicolumn{3}{c}{$<2.4$} \\
\cline{2-4}
\multicolumn{1}{c}{} & 0 \cPqb-tag & 1 \cPqb-tag  & 2 \cPqb-tags \\
\hline
\cPqb-tag & None & One loose &  Medium\,\&\,loose\\
Ang. LD  & $> 0.28$ & $> 0.35$ & $> 0.21$   \\
$QG$ discr. &  $>0.10$ & none & none  \\
$2\ln{\lambda} (\MET)$  &  none & none & $<10$ \\
\hline
${ \mjj}$  &  \multicolumn{3}{c}{ [75, 105]\GeVnn} \\
${ \mll}$   & \multicolumn{3}{c}{ [70, 110]\GeVnn} \\
${ \mZZ}$  &  \multicolumn{3}{c}{ [375, 1500]\GeVnn}\\
\hline
\end{tabular}
\label{table-selection}
\end{center}
\end{table}

\subsection{Summary of selection}
\label{sec:sumsel}

The selection requirements are summarized in
Table~\ref{table-selection}.  When an event contains multiple $\Grav$
candidates passing these selection requirements, the one with the
largest number of b-tagged jets is retained, thus giving preference
to the purest category.  Further ambiguity between multiple candidates
is resolved by selecting the candidate with $\mjj$ (before the kinematic fit) and
$\mll$ values closest to the $\Zo$ boson mass
\mZ. Signal selection efficiencies depend on the graviton mass considered
and range approximately from 3 to 10\% (3 to 21\%) for the RS1 (ADPS) model.

\section{Event analysis}
\label{sec:analysis}

\subsection{Background estimation}
\label{sec:bkgd}

In order to minimize systematic uncertainties associated with the
background shapes, we
estimate the background distribution from the data.  The $\mZZ$
distribution of the selected events is split into three samples
based on the number of \cPqb-tagged jets.
The selected events are then examined comparing to 33 equally-spaced
hypothetical
graviton masses in the range between 400\GeVnn\ and 1200\GeVnn.

A signal window in the $\mZZ$ distribution is defined around the
expected position of the signal peak. The signal shape is parameterized
using the sum of a double Crystal-Ball function (i.e., a Gaussian
distribution
with power-law tails on both sides)~\cite{CrystalBall} and an
empirical line-shape
derived from the
observed distribution of events with misassigned jets (a triangular shape
convoluted with a Crystal-Ball function). The latter component
describes
both mismeasurements and incorrect reconstruction of the graviton decay
products. The size of the signal window corresponds to three times
the standard deviation of the core distribution of the signal shape, which
ranges from 13 (at $M_1 = 400\GeV$) to 38\GeVnn (at $M_1 = 1200\GeV$). For
every mass value considered, the $\mZZ$ distribution is fitted in the
$[375, 1500]\GeV$ range after excluding the signal window, in order to
estimate the background expectation from the sideband regions.
The distributions of
$\mZZ$ for the background are parameterized with exponential
functions. Goodness of fit is estimated using binned $\chi^2$ tests.
To gain stability in the fits, we combine the electron and muon
samples, assuming that the background shapes are the same: it has been verified
that the fitted slopes for separate samples agree within statistical
uncertainties.

\begin{figure*}[htp]
\begin{center}
\includegraphics[width=0.38\textwidth]{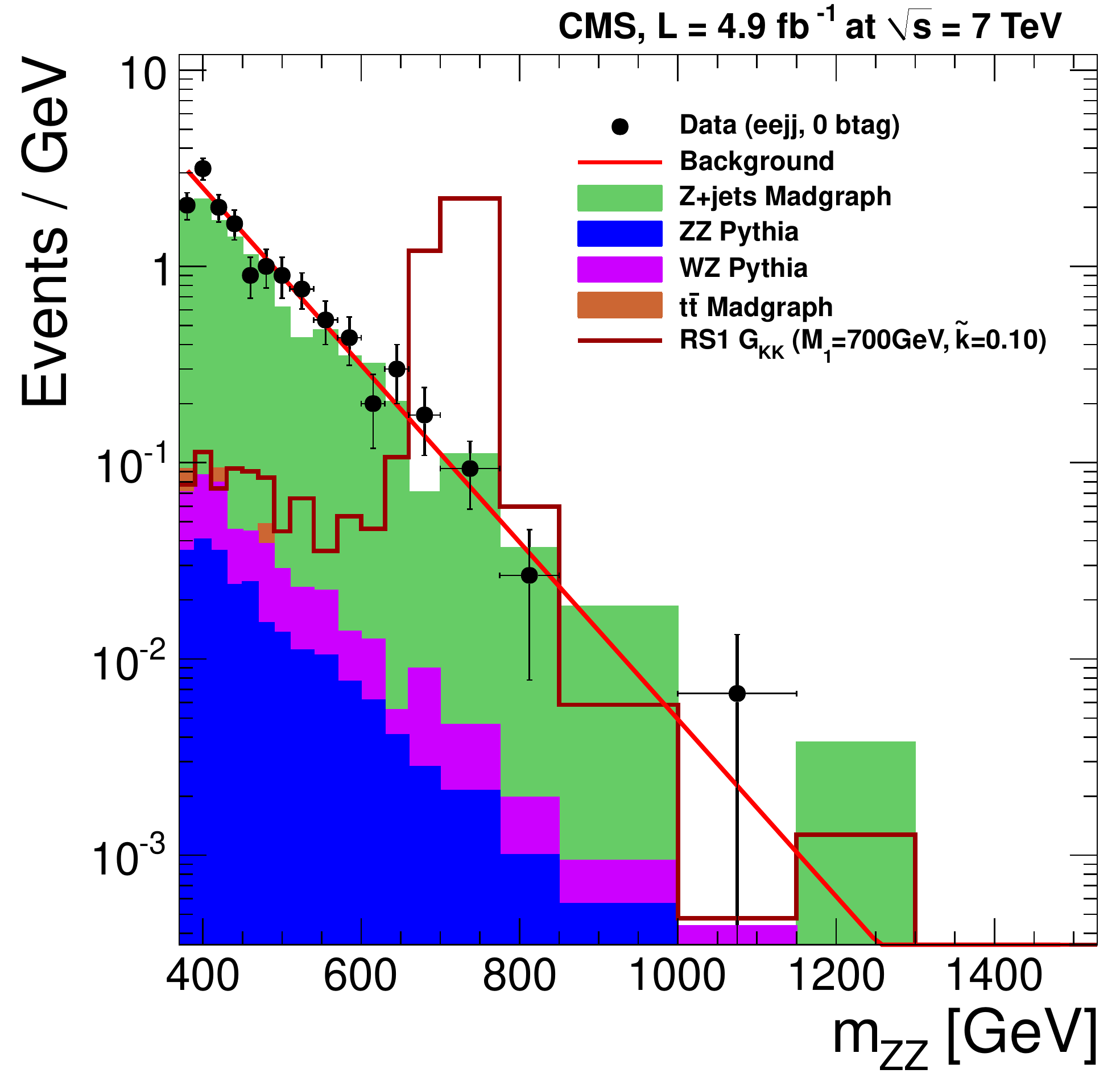}
\includegraphics[width=0.38\textwidth]{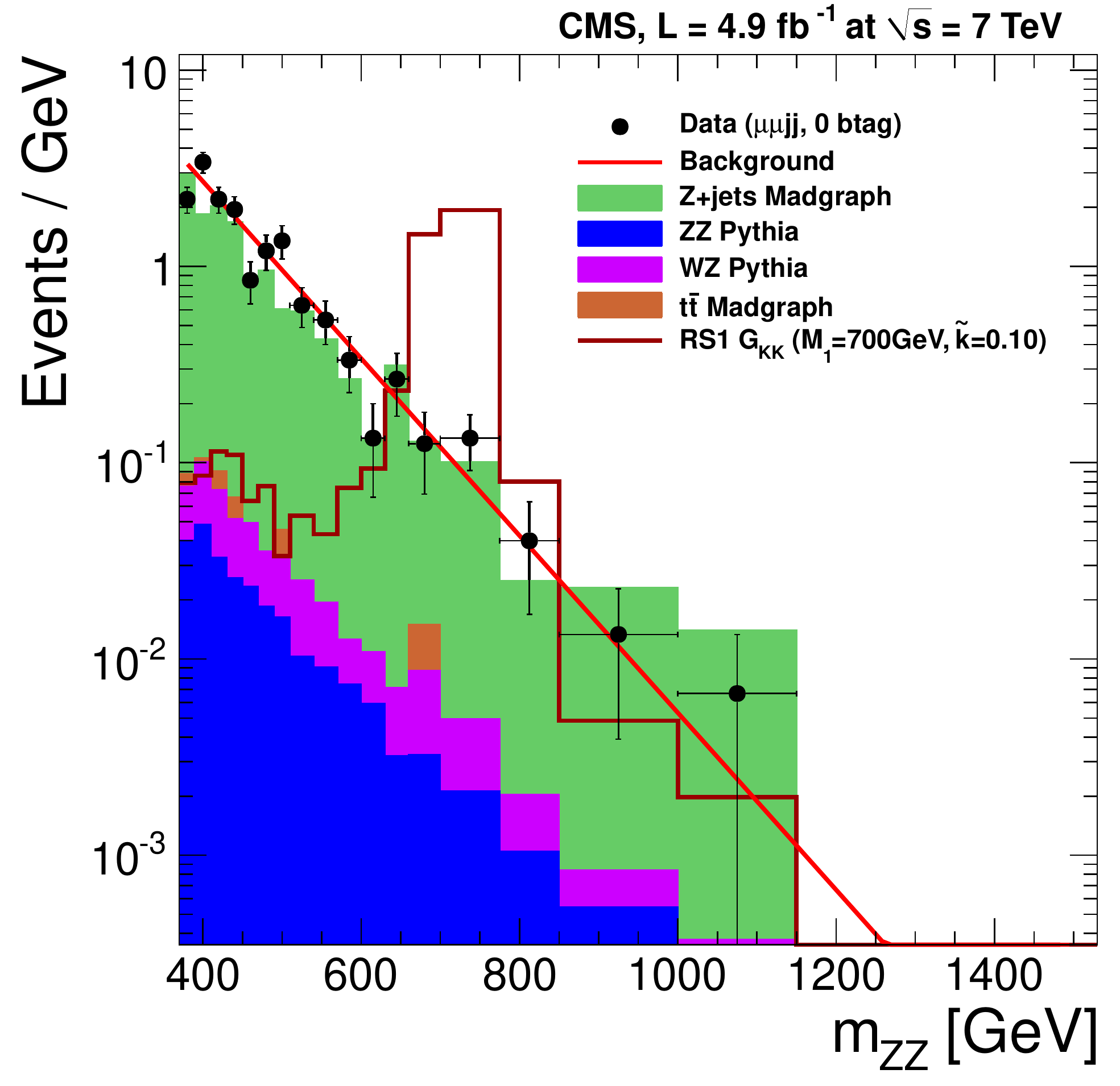}\\
\includegraphics[width=0.38\textwidth]{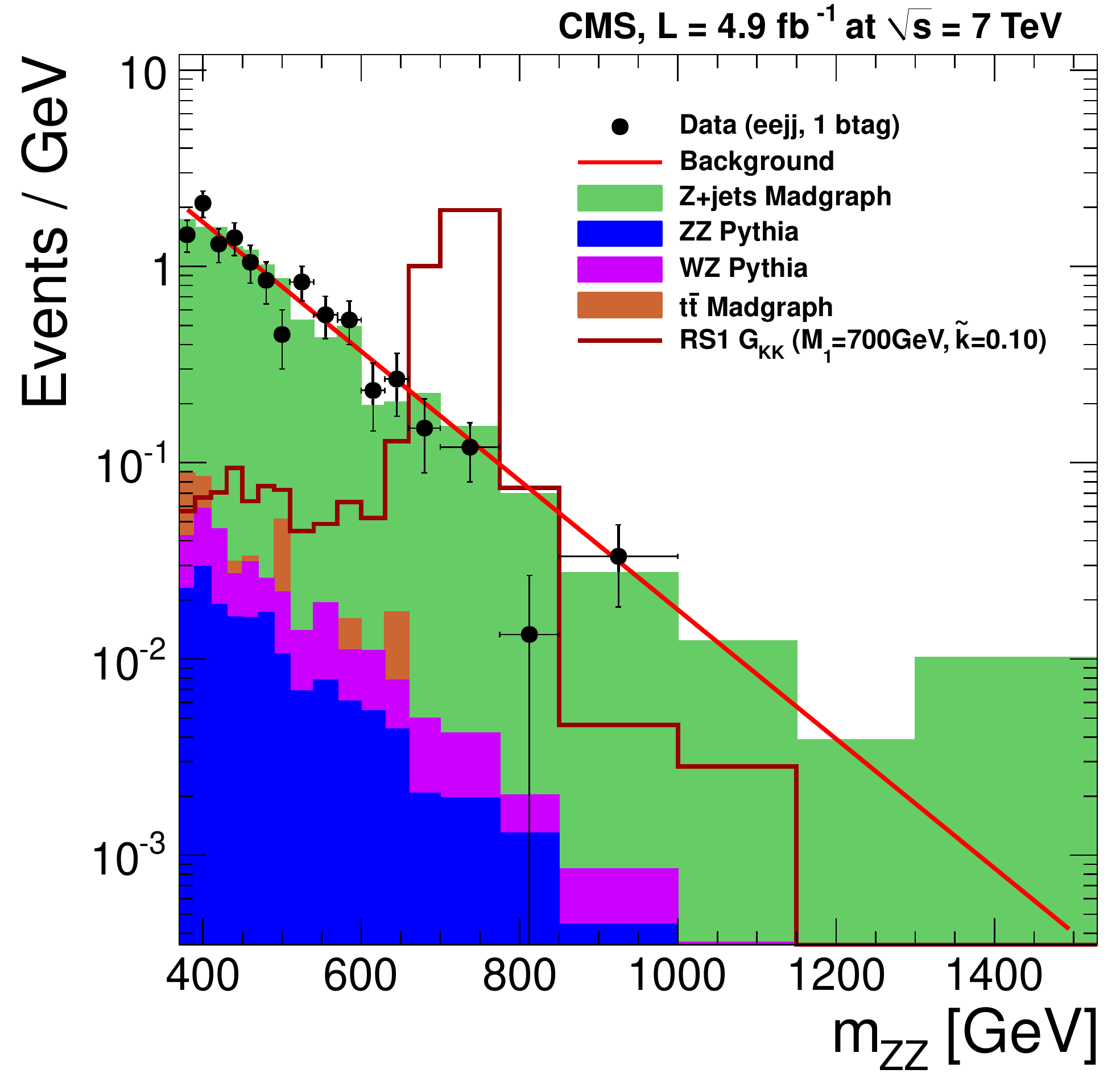}
\includegraphics[width=0.38\textwidth]{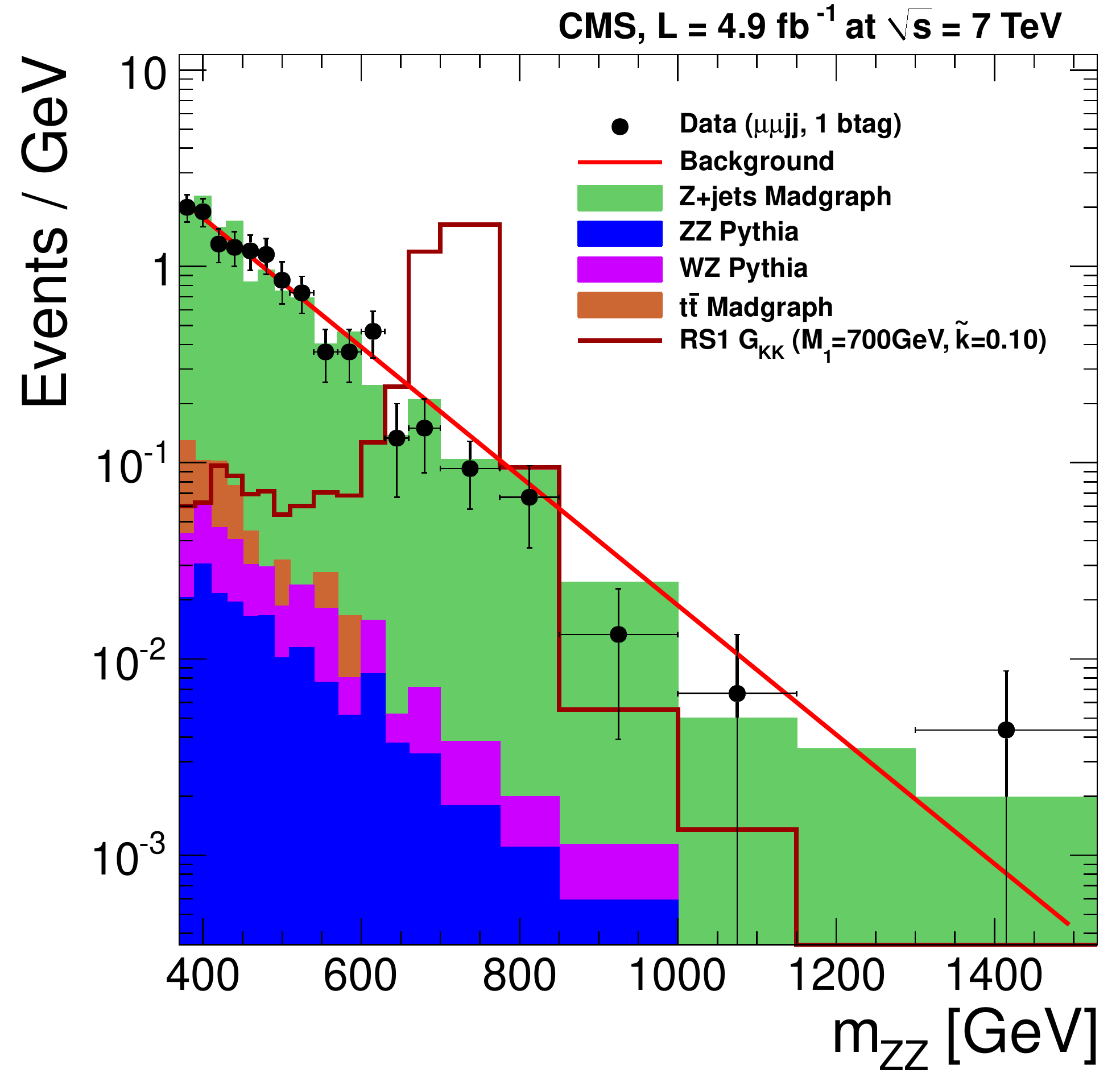} \\
\includegraphics[width=0.38\textwidth]{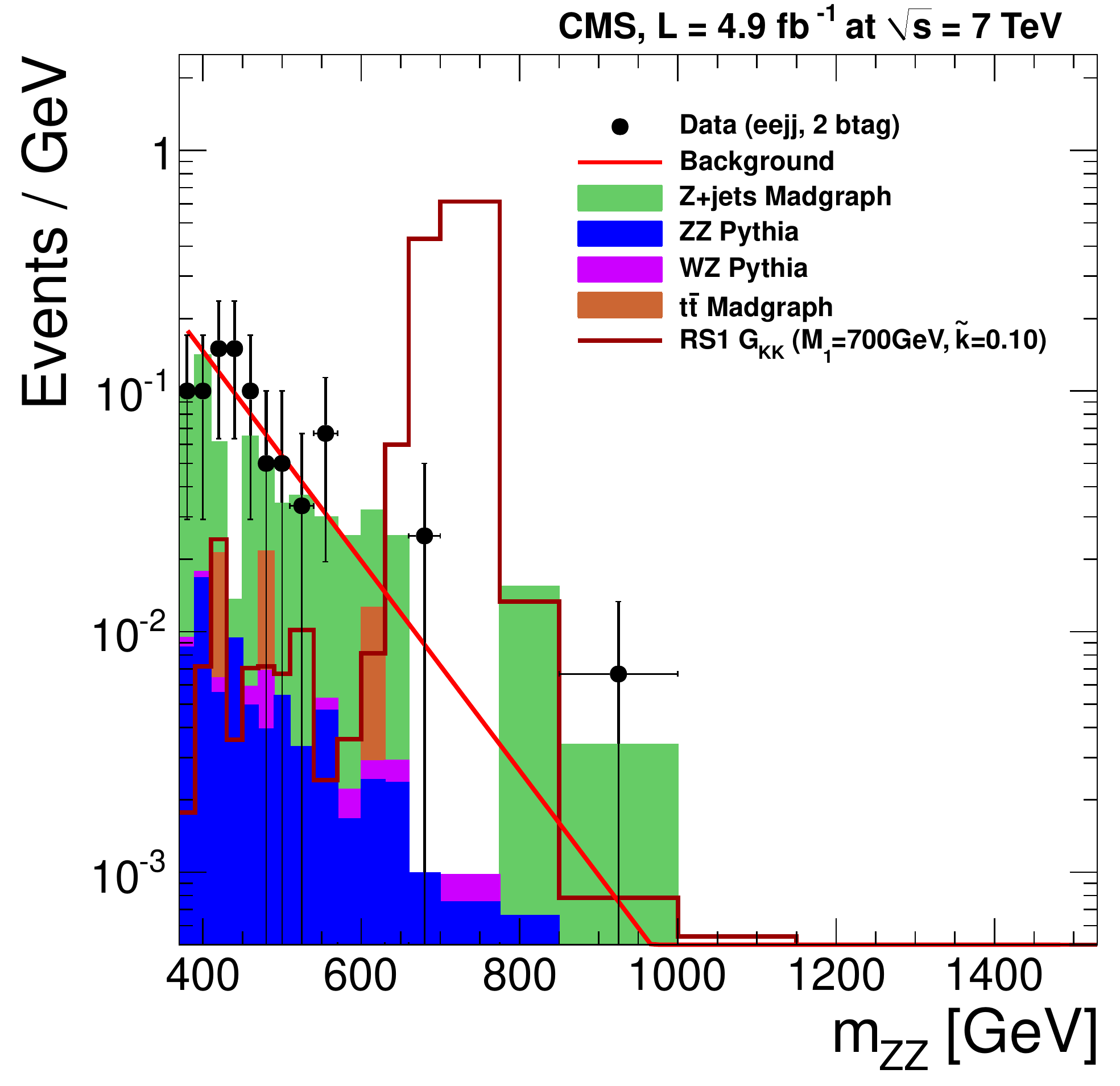}
\includegraphics[width=0.38\textwidth]{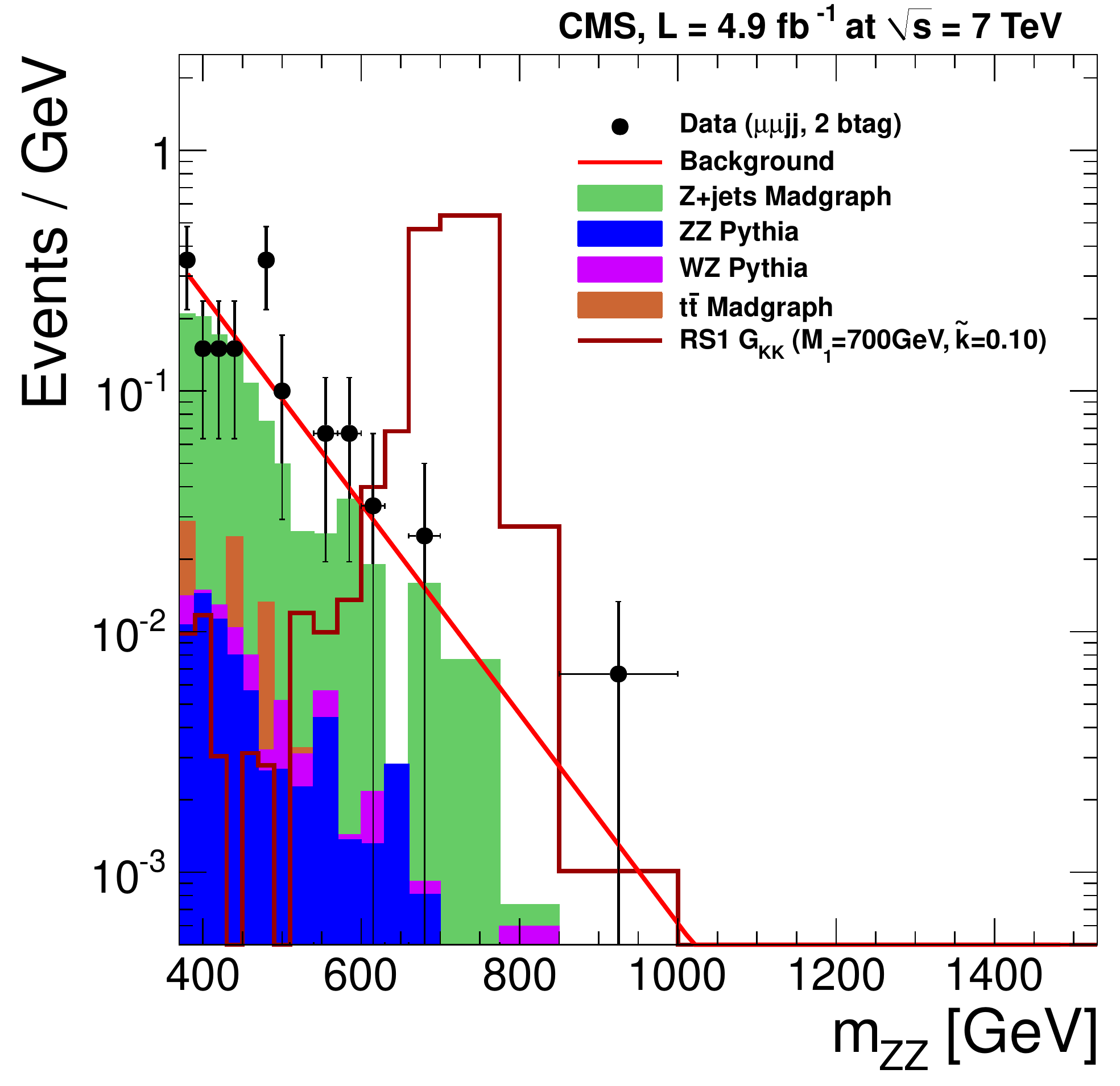}
\caption{
The $m_{\cPZ\cPZ}$ invariant mass distribution after final selection in
three categories: 0~b-tag (top), 1~b-tag (middle), 2~b-tag
(bottom). The two channels $\Pe\Pe jj$ and $\Pgm\Pgm jj$ are
presented separately on the left and on the right,
respectively. Points with error bars show distributions of data,
stacked filled histograms show the background expectation from
simulated events with the different components illustrated.  Also
shown is a hypothetical RS1 graviton signal with mass of 700\GeV and
$\ktilde=0.10$.  The red (lighter) solid line shows the central value
of the prediction of background from the sideband extrapolation
procedure (excluding a signal window of ${\pm}57\GeV$ around 700\GeV
in this example). }
\label{fig:mZZ_kinfit_hiMass}
\end{center}
\end{figure*}

Figure~\ref{fig:mZZ_kinfit_hiMass} shows the result of the fit for the
specific case of $\mZZ=700\GeV$, overlaid with the $\mZZ$ distributions
split by both \cPqb-tag category and leptonic channel. The normalization of the fit is determined independently for each
plot.
The MC prediction normalized to the integrated luminosity of the data sample is also included
in the plot. Although the agreement of the MC prediction with the data is fair,
differences are observed in both normalization and shape.
The most significant discrepancy between data and MC,
in the 0 \cPqb-tag category (11\% in the total yield), is because
the relative fraction of quark-
and gluon-originated jets is not well reproduced in \MADGRAPH MC, which affects the $QG$ selection
efficiency. Because the background is determined directly from the
sidebands in the data, the results are not
sensitive to the limitations of the simulated samples.

The yields
 predicted by
the background estimation from control samples in data  are
 compared to the observed ones in
 Table~\ref{table:ExpectedYields}.
Additionally, the expected number
 of events for the hypothetical graviton signals in the RS1 and ADPS
 models are listed for some benchmark mass values.

\begin{table}[hbt]
\begin{center}
\topcaption{
Observed and expected yields in the 4.9\fbinv dataset at
$\sqrt{s}=7\TeV$. The yields
are quoted in the range $375 <m_{\cPZ\cPZ}< 1500\GeV$ (excluding a signal window of ${\pm}57\GeV$
around 700 \GeV in this example).
The signal expectations for the RS1 and ADPS
models are reported for three benchmark $M_1$ points. The uncertainties shown
are the sum in quadrature of the systematic and statistical
uncertainties.}
\begin{tabular}[\linewidth]{lc|c|c}
 \multicolumn{1}{c}{}   & 0 b-tag & 1 b-tag  & 2 b-tags \\
\hline
{Obs. yield}     & 658  & 542 & 51 \\
{Exp. bkgnd}   & 652$\pm$25  & 548$\pm$23  & 51$\pm$7 \\
\multicolumn{2}{l|}{(700\GeV signal hyp.)} & & \\
\hline \hline
 \multicolumn{4}{c}{ } \\[-2ex]
 \multicolumn{4}{c}{ Signal expectation (RS1 graviton, \ktilde=0.10) }   \\  [0.5ex]
 \hline
 500\GeV        & 268$\pm$24    & 212$\pm$16     & 100$\pm$12 \\
 700\GeV        & 50$\pm$5  & 39$\pm$4   & 12.4$\pm$1.6 \\
 900\GeV        & 8.4$\pm$1.2  & 7.2$\pm$0.8   & 1.36$\pm$0.24 \\
\hline \hline
 \multicolumn{4}{c}{ } \\[-2ex]
 \multicolumn{4}{c}{ Signal expectation (ADPS graviton, \ktilde=0.50) }   \\   [0.5ex]
 \hline
 500\GeV        &    30.4$\pm$2.5     & 25.8$\pm$1.5     &  10.6$\pm$1.3\\
 700\GeV        &    2.9$\pm$0.3      & 2.8 $\pm$0.2     &  0.81$\pm$0.10\\
 900\GeV        &    0.20$\pm$0.03    & 0.18$\pm$0.01    &  0.025$\pm$0.003\\
\hline
\end{tabular}
\label{table:ExpectedYields}
\end{center}
\end{table}

\subsection{Systematic uncertainties}
\label{sec:sysunc}

The uncertainties on the normalization of the
background fit are statistical and depend on how many
events populate the sidebands (uncertainties reported in Table~\ref{table:ExpectedYields}).
The shape uncertainty is determined by
propagating the uncertainty on the slope parameter of the exponential
into the background estimation. Because of the steep fall of the
$\mZZ$ spectrum, these uncertainties are larger for low values of the
graviton masses. In the case of the search for a $400\GeV$ (700\GeV) resonance, the uncertainties on the shape of
the background are 10\%
(4\%) for the 0 \cPqb-tag, 10\% (5\%) for the 1
\cPqb-tag, and 30\% (15\%) for the 2 \cPqb-tag category.
Alternative functions (inverse power law and exponential with modified tail
shape) have been used for the background fit, and the descriptions obtained
agree with the nominal one within the statistical uncertainties of the slope
parameter. Hence, no additional systematic uncertainty is considered from
this source.

The systematic uncertainties on the signal normalization are summarized in
Table~\ref{table-systematics}.  We consider effects from lepton energy
scale, resolution, selection, and trigger; jet efficiency, energy scale
and resolution
after calibration; pileup; $\MET$ significance requirements;
heavy-quark flavor tagging and quark-gluon discrimination; graviton
production mechanism; and the luminosity measurement.

Reconstruction efficiencies for leptons and their uncertainties are
evaluated from control samples in data~\cite{VBTF}.
The systematic uncertainties on jet reconstruction are evaluated by
variation of the jet energy scale and resolution within calibration
uncertainties. Uncertainties due to pileup are taken
as the difference
between reconstruction and selection efficiencies with the number of pileup
vertices shifted by one unit up and down with respect to the average
value measured in data.  The requirement on the discriminant
$\lambda(\MET)$
translates into 1--2\% reduction in signal and the resulting
uncertainty is taken as the full inefficiency value. Uncertainties on the
quark-gluon selection efficiency are evaluated using a selected
$\gamma$+jet sample in data, which predominantly contains quark
jets. The uncertainty on the
$\cPqb$ tagging has several sources, such as pileup effects and dependence
on the fraction of b production from gluon splitting,
and has been evaluated on a sample of dijet events
enriched in heavy-flavor by requiring a muon to be associated with
one jet~\cite{CMS-PAS-BTV-11-001}.

For a given production model, there is an additional uncertainty in the
production mechanism coming from the choice of PDFs.
PDFs affect the distribution of the longitudinal momentum of the graviton,
hence the signal acceptance.  We follow the
PDF4LHC~\cite{Botje:2011sn,Alekhin:2011sk,Lai:2010vv,Martin:2009iq,Ball:2011mu}
recommendation to estimate the uncertainty due to PDF knowledge and
to calculate the uncertainty on signal acceptance.
The relative
uncertainty on the integrated luminosity~\cite{CMS-PAS-SMP-12-008} is also applied to the signal normalization.

\begin{table}[t]
\begin{center}
\topcaption{
Summary of sources of systematic uncertainties on signal normalization. All
are expressed as relative uncertainties on the cross section.
}
\begin{tabular}{lccc}
 Source      &   0 \cPqb-tag   &   1 \cPqb-tag  &   2 \cPqb-tags \\
\hline
\Pgm\  reconst. \& trig.   &  \multicolumn{3}{c}{2.7\%}   \\
\Pe\ reconst. \& trig.  &  \multicolumn{3}{c}{4.5\%}     \\
Jet reconst.      &  \multicolumn{3}{c}{2--4\%}\\
Pileup              &  \multicolumn{3}{c}{1--4\%}    \\
$\MET$ significance                &  -- & -- & 1--2\%  \\
$\cPqb$ tagging    &  2--7\% & 3--5\% & 10--11\%  \\
$QG$-discrimination    & 4.6\%   & -- & -- \\
Acceptance (PDF) &  \multicolumn{3}{c}{$\sim$1\%}  \\
Luminosity            &  \multicolumn{3}{c}{2.2$\%$}   \\
\hline
\end{tabular}
\label{table-systematics}
\end{center}
\end{table}

\section{Results}
\label{sec:results}

Since no significant excess is observed in the data distributions, we set
exclusion limits at the 95\% confidence level (CL) on the product of the
graviton cross section and the branching fraction of $\GZZ$.
The upper limits are calculated using the $\mathrm{CL_s}$
\cite{LEP-CLs2, LEP-CLs} method.

Based on the normalization and shape of the $m_{\cPZ\cPZ}$
distribution for signal and background, obtained as described in
Section~\ref{sec:bkgd}, we perform an unbinned
statistical analysis of the results using the same formalism as in
\cite{HIG-11-027, HIG-11-032}. The six data categories are combined in
an unbinned maximum likelihood fit for each mass hypothesis with a
background-only and a signal-plus-background model, and the likelihood ratio
of the two fits serves as test statistic for the $\mathrm{CL_s}$ procedure.
Systematic uncertainties are incorporated as nuisance parameters and are treated
according to the frequentist paradigm.

\begin{figure}[hbpt]
\centering
\includegraphics[width=0.48\textwidth]{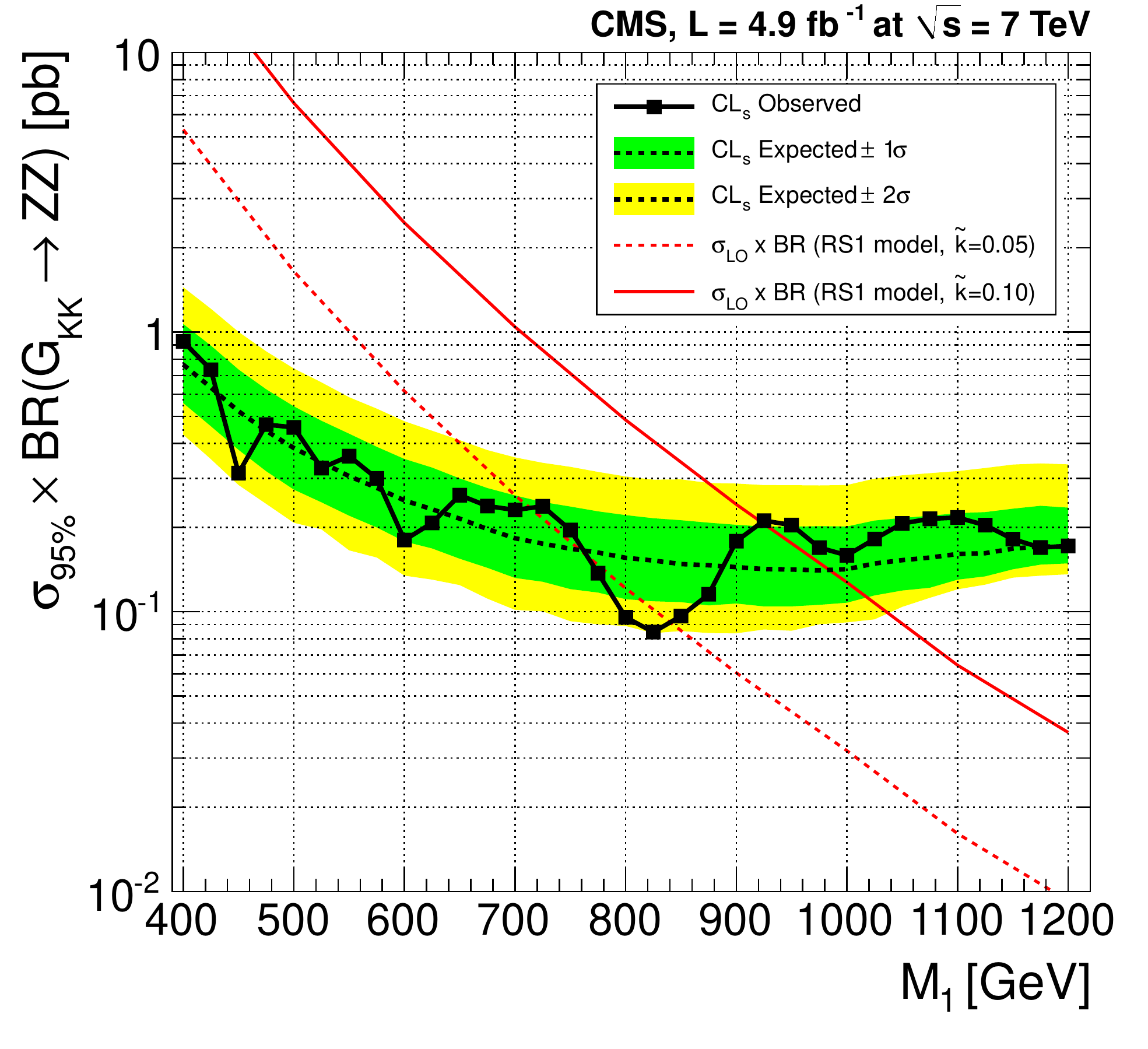}
\includegraphics[width=0.48\textwidth]{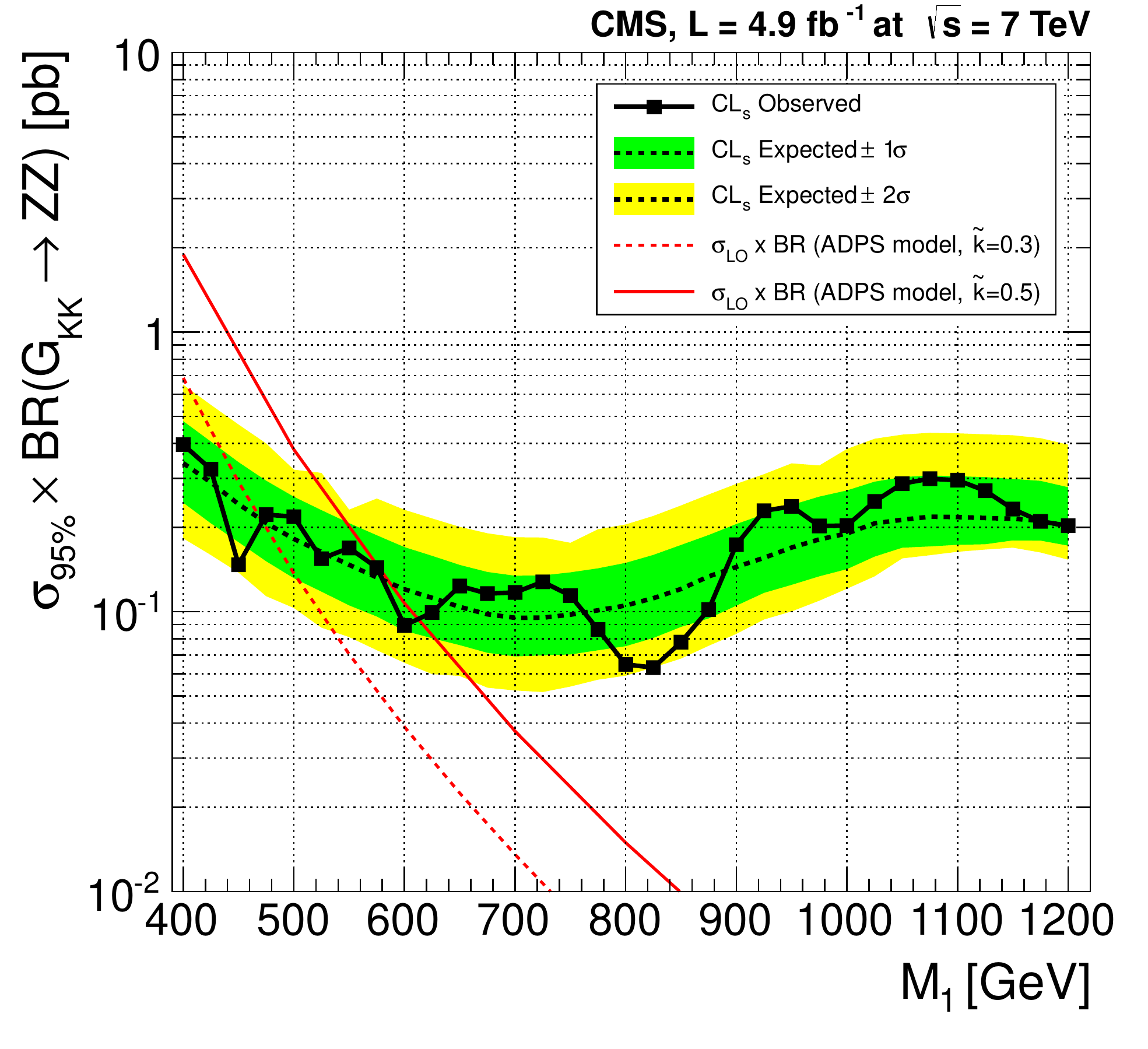}
\caption{
Observed (black solid lines) and expected (black dashed line) 95\% CL upper
limits on the cross section times branching fraction for $\GZZ$,
obtained with the $\mathrm{CL_s}$ technique. The graviton signal was
generated according to the model in \cite{RS1Graviton1}
(RS1, \cmsLeft) and
\cite{RSGravAgashe1} (ADPS, \cmsRight).  The 68\%
and 95\% ranges of expectation for the background-only model are also
shown with green (darker) and yellow (lighter) bands surrounding
the expected upper limit line, respectively.
The predicted product
of the graviton cross section and the branching fraction for different values
of $\ktilde$ are shown as red solid and dashed curves.  }
\label{fig:plot_sigmaAgashe}
\end{figure}

The resulting upper limits are compared to the theoretical cross
sections times branching fraction of $\GZZ$.
Figure~\ref{fig:plot_sigmaAgashe} (\cmsLeft) shows the upper limits
for the RS1 model and compares them to LO theoretical expectations for
two benchmark values of the coupling strength $\ktilde$, equal to 0.05
and 0.10. These results allow us to exclude at the 95\% CL gravitons
with $M_{1} < 945\GeV$ for $\ktilde = 0.10$, improving on previous
analyses~\cite{ATLAS:ZZsearch} where a phase-space decay model of the
graviton is used. For a value of the coupling strength $\ktilde =
0.05$, we exclude the separate ranges $M_{1} < 720\GeV$ and $760 <
M_{1} < 850\GeV$, because of an under-fluctuation of the observed
limit around $M_{1} = 800\GeV$.  The upper limit for a signal
generated according to the ADPS model is shown in
Fig.~\ref{fig:plot_sigmaAgashe} (\cmsRight). The limit on the
production cross sections times branching fraction of $\GZZ$, as a
function of the graviton mass, is compared to the model predictions at
LO, resulting in an exclusion range of $M_{1} < 610\GeV$ for $\ktilde =
0.50$.

\section{Summary}
\label{sec:concl}

Results have been presented of a search for a KK graviton
decaying into ZZ, with one Z-boson decaying leptonically, the other
hadronically.
CMS data at $\sqrt{s}=7\TeV$ have been used,
corresponding to a total integrated luminosity of 4.9\fbinv.
As a novel feature of this study,
the analysis targets specific graviton models, by exploiting information
on angular distributions of the decay products to build likelihood-ratio
discriminants that enhance the signal sensitivity.

No excesses in the $\mZZ$ distributions over the expected SM
backgrounds are found, for the range
400--1200\GeV. Upper limits on the production cross section times
branching fraction as a function of graviton mass are set
at 95\% confidence level.
Exclusion ranges for the RS1 model are
$M_{1} < 945\GeV$ for $\ktilde = 0.10$, and $M_{1}
< 720\GeV$ and $760 < M_{1} < 850\GeV$ for $\ktilde = 0.05$.

For the first time, upper limits are obtained specifically for the
ADPS model, which is more compatible with current indirect
experimental limits than the original RS model. We place an exclusion
range of $M_{1} < 610\GeV$ for $\ktilde = 0.50$.

\section*{Acknowledgements}
We congratulate our colleagues in the CERN accelerator departments for the excellent performance of the LHC machine. We thank the technical and administrative staff at CERN and other CMS institutes, and acknowledge support from: BMWF and FWF (Austria); FNRS and FWO (Belgium); CNPq, CAPES, FAPERJ, and FAPESP (Brazil); MES (Bulgaria); CERN; CAS, MoST, and NSFC (China); COLCIENCIAS (Colombia); MSES (Croatia); RPF (Cyprus); MoER, SF0690030s09 and ERDF (Estonia); Academy of Finland, MEC, and HIP (Finland); CEA and CNRS/IN2P3 (France); BMBF, DFG, and HGF (Germany); GSRT (Greece); OTKA and NKTH (Hungary); DAE and DST (India); IPM (Iran); SFI (Ireland); INFN (Italy); NRF and WCU (Korea); LAS (Lithuania); CINVESTAV, CONACYT, SEP, and UASLP-FAI (Mexico); MSI (New Zealand); PAEC (Pakistan); MSHE and NSC (Poland); FCT (Portugal); JINR (Armenia, Belarus, Georgia, Ukraine, Uzbekistan); MON, RosAtom, RAS and RFBR (Russia); MSTD (Serbia); SEIDI and CPAN (Spain); Swiss Funding Agencies (Switzerland); NSC (Taipei); TUBITAK and TAEK (Turkey); STFC (United Kingdom); DOE and NSF (USA). Individuals have received support from the Marie-Curie programme and the European Research Council (European Union); the Leventis Foundation; the A. P. Sloan Foundation; the Alexander von Humboldt Foundation; the Austrian Science Fund (FWF); the Belgian Federal Science Policy Office; the Fonds pour la Formation \`a la Recherche dans l'Industrie et dans l'Agriculture (FRIA-Belgium); the Agentschap voor Innovatie door Wetenschap en Technologie (IWT-Belgium); the Council of Science and Industrial Research, India; the Compagnia di San Paolo (Torino); and the HOMING PLUS programme of Foundation for Polish Science, cofinanced from European Union, Regional Development Fund.

\bibliography{auto_generated}   

\cleardoublepage \appendix\section{The CMS Collaboration \label{app:collab}}\begin{sloppypar}\hyphenpenalty=5000\widowpenalty=500\clubpenalty=5000\textbf{Yerevan Physics Institute,  Yerevan,  Armenia}\\*[0pt]
S.~Chatrchyan, V.~Khachatryan, A.M.~Sirunyan, A.~Tumasyan
\vskip\cmsinstskip
\textbf{Institut f\"{u}r Hochenergiephysik der OeAW,  Wien,  Austria}\\*[0pt]
W.~Adam, E.~Aguilo, T.~Bergauer, M.~Dragicevic, J.~Er\"{o}, C.~Fabjan\cmsAuthorMark{1}, M.~Friedl, R.~Fr\"{u}hwirth\cmsAuthorMark{1}, V.M.~Ghete, J.~Hammer, N.~H\"{o}rmann, J.~Hrubec, M.~Jeitler\cmsAuthorMark{1}, W.~Kiesenhofer, V.~Kn\"{u}nz, M.~Krammer\cmsAuthorMark{1}, I.~Kr\"{a}tschmer, D.~Liko, I.~Mikulec, M.~Pernicka$^{\textrm{\dag}}$, B.~Rahbaran, C.~Rohringer, H.~Rohringer, R.~Sch\"{o}fbeck, J.~Strauss, A.~Taurok, W.~Waltenberger, G.~Walzel, E.~Widl, C.-E.~Wulz\cmsAuthorMark{1}
% ...
\vskip\cmsinstskip
\dag:~Deceased\\
1:~~Also at Vienna University of Technology, Vienna, Austria\\
% ...

\end{sloppypar}
\end{document}